\documentstyle [aps]{revtex}
\newcommand{\eref}[1]{(\ref{#1})}
\newcommand{\mat}[1]{\underline{\underline{\bf #1}}}
\newcommand{\PR}{{\it Phys. Rev. }}
\newcommand{\PRL}{{\it Phys. Rev. Lett. }}
\newcommand{\PL}{{\it Phys. Lett. }}

\begin{document}
\title{Enhancement of the electric dipole moment of the
electron in the YbF molecule\footnote{To be submitted to J.~Phys.~B}}
\author{M G Kozlov}
\address{Petersburg Nuclear Physics Institute,
Gatchina, Leningrad district, 188350, Russia}
\maketitle

\begin{abstract}
We calculate an effective electric field on the unpaired electron in
the YbF molecule. This field determines sensitivity of the molecular
experiment to the electric dipole moment of the electron. We use
experimental value of the spin-doubling constant $\gamma$ to estimate
the admixture of the configuration with the hole in the 4f-shell of
Ytterbium to the ground state of the molecule. This admixture
reduces the field by $7$\%.  Our value
for the effictive field is 5.1 a.u. = $2.5 \cdot 10^{10}$~V/cm.
\end{abstract}


\subsection*{Introduction}

It is well known that effects caused by an electric dipole moment ({\sc
EDM}) of the electron $d_{\rm e}$ are strongly enhanced in heavy diatomic
radicals (see, for example, a review \cite{KL}). An experimental search for
the {\sc EDM} of the electron is now underway in Brighton University on the
YbF molecule \cite{SPRH}. For this reason the reliable calculations for
this molecule are necessary. The existence of a shallow 4f-shell adds
complexity to such calculations. Here we extend the semiempirical approach
suggested in \cite{K85,KE} to account for a possible admixture of the
configuration with the hole in the 4f-shell of Ytterbium. The idea that
f-hole can explain small value of the spin-doubling constant $\gamma$
belongs to I.~B.~Khriplovich \cite{IBK}.

The {\sc EDM} experiments with radicals are made on the spin-rotational
levels of the electronic ground state. The spin-rotational degrees of
freedom of the molecule are described by the following spin-rotational
Hamiltonian \cite{SF,GLM,FK,KLM}:
\begin{eqnarray}
        H_{\rm sr} & = & B{\bf N}^2 + \gamma {\bf S N}
                 + {\bf S \mat{A} I}
                 + W_d d_{\rm e}
                 {\bf S n}.
\label{1}
\end{eqnarray}
In this expression
$\bf N$ is the rotational angular momentum, $B$ is the rotational constant,
$\bf S$ is the spin of the electron and
$\bf I$ is the spin of the Yb nucleus,
$\bf n$ is the unit vector directed along the molecular axis from Yb to F.
The spin-doubling constant $\gamma$ characterizes the spin-rotational
interaction. The axial tensor $\mat{A}$ describes magnetic
hyperfine structure. It can be determined by two parameters:
$A=(A_{\parallel}+2A_{\perp})/3$ and $A_{\rm
d}=(A_{\parallel}-A_{\perp})/3$. The last term
in \eref{1} correspond to the interaction of the {\sc EDM} of the
electron $d_{\rm e}$ with the molecular field, $\case{1}{2}W_d$ being
the effective electric field on the electron.

Parameters $B$, $\gamma$, $A$ and $A_{\rm d}$ are known from the experiment
\cite{HH,SWH,KW}:
\begin{eqnarray}
        B = 7237 \mbox{ MHz},\quad \gamma = 13 \mbox{ MHz}, \quad
        A = 7617 \mbox{ MHz}, \quad A_{\rm d} = 102 \mbox{ MHz},
\label{1a}
\end{eqnarray}
while $W_d$ have to be calculated. There are three calculations of
this parameter \cite{KE,TME,Q}. In earlier calculations \cite{KE,TME}
4f-shell of Ytterbium was frozen. In \cite{Q} f-electrons are included
in the valence space, but details of this calculation are not yet
published. Here we allow the admixture of the f-hole to the ground
state of the molecule.

\subsection*{Electronic wave function}

The ground state of YbF molecule is known to be $\Sigma_{1/2}$ \cite{HH}.
The large hyperfine constants $A$ and $A_{\rm d}$ indicate that unpaired
electron occupies molecular orbital $\sigma_{\rm s}$ with dominant
contribution from 6s-orbital and significant contribution of 6p$_0$-orbital
of Yb ion. But a simple one-configurational wave function can not explain
an exceptionally small value of the constant $\gamma$.  Following
Khriplovich we will assume that there is small admixture of the f-hole:
\begin{eqnarray}
        |\Sigma,\omega \rangle
        = |\sigma_{\rm s},\omega \rangle
        + \delta_{\rm f} \overline{|\sigma_{\rm f},-\omega \rangle},
\label{2}
\end{eqnarray}
where $\omega=\pm \case12$ is the projection of the total electronic angular
momentum on the molecular axis and the bar over the orbital corresponds to
the hole.  More explicitly \eref{2} means that many electron wave
function reads
\begin{eqnarray}
        |\Sigma,\omega \rangle = [\dots] \left(
        |\sigma_{\rm f},-\case12 \rangle |\sigma_{\rm f},\case12 \rangle
        |\sigma_{\rm s},\omega \rangle + \delta_{\rm f}
        |\sigma_{\rm f},\omega \rangle |\sigma_{\rm s},-\case12 \rangle
        |\sigma_{\rm s},\case12 \rangle \right),
\label{3}
\end{eqnarray}
where $[\dots]$ denotes the closed core.  The spin-orbit interaction for
the f-hole is large and has the opposite sign. This can explain the
small value of $\gamma$ (see below).

All constants of the Hamiltonian \eref{1} except for the
constant $B$ depend only on the electron spin density in the vicinity of
the heavy nucleus. For this reason we can expand molecular orbitals
in spherical waves with the origin at the Yb nucleus:
\begin{eqnarray}
        |\sigma_{\rm s},\omega \rangle
        &=&(x_{\rm s}|\tilde{6}s \rangle + x_{\rm p} |\tilde{6}p_0 \rangle
        + x_{\rm d} |\tilde{5}d_0 \rangle + \dots)|\omega \rangle,
\label{4} \\
        |\sigma_{\rm f},\omega \rangle &=& (|\tilde{4}f_0 \rangle
        +\varepsilon_{\rm d}|\tilde{5}d_0 + \dots \rangle) |\omega \rangle,
\label{5}
\end{eqnarray}
where $|\omega \rangle$ denotes spin function and dots stand for the
higher spherical waves. The radial functions in this expansion can be
considered as distorted orbitals of the Yb$^+$ ion. Corresponding principle
quantum numbers are marked with tilde.

Wave function \eref{2} is written in a pure nonrelativistic coupling case.
Spin-orbit interaction $H_{\rm so}$ mixes state \eref{2} with $\Pi_{1/2}$
states. One can expect that the largest admixtures correspond to the
molecular orbitals $\pi_{\rm p}$ and $\pi_{\rm f}$ (the spin-orbit
interaction for the 5d-shell is much smaller):
\begin{eqnarray}
        |\pi_{\rm p},\omega \rangle &=& (a_{\rm p} |\tilde{6}p_{2\omega}
        \rangle + \dots)|-\omega \rangle,
\label{6} \\
        |\pi_{\rm f},\omega \rangle
        &=& (|\tilde{4}f_{2\omega} \rangle + \dots)|-\omega \rangle.
\label{7}
\end{eqnarray}
Then, the wave function of the ground state has the form
\begin{eqnarray}
        |\Sigma,\omega \rangle
        = |\sigma_{\rm s},\omega \rangle
        + c_{\rm p} |\pi_{\rm p},\omega \rangle
        + \delta_{\rm f} \overline{|\sigma_{\rm f},-\omega \rangle}
        + c_{\rm f} \overline{|\pi_{\rm f},-\omega \rangle}.
\label{7a}
\end{eqnarray}

\subsection*{Hyperfine tensor and parameter $W_d$}

The operator of the hyperfine interaction in atomic units has the form
\begin{eqnarray}
     H_{\rm hf} = \frac{g_{\rm n} \alpha}{2m_{\rm p}}
     (\vec{\alpha} \times {\bf r} \cdot{\bf I})\frac{1}{r^3},
\label{17a}
\end{eqnarray}
where $g_{\rm n}$ is the nuclear $g$-factor, $\alpha$ is the fine structure
constant, $m_{\rm p}$ is the proton mass and $\vec{\alpha}$ is the vector of
Dirac matrices. This operator is known to be almost diagonal in quantum
number $l$ and in the following calculations we neglect nondiagonal terms.

Let us start with the nonrelativistic expressions for a spherical wave
$l \neq 0$. It is not difficult to derive, that for the $\sigma$-type state
$|l,0 \rangle$
\begin{eqnarray}
     A\, = 0, \qquad
     A_{\rm d} = \frac{g_{\rm n} \alpha^2}{2m_{\rm p}}
             \frac{l(l+1)}{(2l-1)(2l+3)}
             \langle n,l|\frac{1}{r^3}|n,l \rangle.
\label{17c}
\end{eqnarray}
So, in the nonrelativistic approximation only s-wave contributes to the
isotropic constant $A$. Note that the radial integral in \eref{17c}
is rapidly decreasing with $l$, while the coefficient in front of it is a
weak function of $l$.

Interaction of the {\sc EDM} of the electron with the molecular electric
field $-\nabla \phi$ is also singular at the nucleus \cite{PGHS}. The most
convenient form of this operator is \cite{IBK1}:
\begin{eqnarray}
        H_d = 2 d_{\rm e}
        \left(
        \begin{array}{cc}
        0 & 0 \\
        0 & \vec{\sigma}
        \end{array} \right)
        (-\nabla \phi).
\label{hs0}
\end{eqnarray}

Relativistic expressions for $A$, $A_{\rm d}$ and $W_d$ can
be found in \cite{K85,KE,KL}.  In the paper \cite{KE} these constants were
calculated for the wave function \eref{7a} with $c_{\rm p}=\delta_{\rm
f}=c_{\rm f}=0$:
\begin{eqnarray}
        A &=& \left(11850 x_{\rm s}^2 -102 x_{\rm p}^2 -5 x_{\rm d}^2
        +\cdots \right) \mbox{ MHz},
\label{hs1}\\
        A_{\rm d} &=& \left(422 x_{\rm p}^2 + 36 x_{\rm d}^2
        +\cdots \right) \mbox{ MHz},
\label{hs2}\\
        W_d &=& \left(-29.7 x_{\rm s} x_{\rm p} + 1.3 x_{\rm p} x_{\rm d}
        +\cdots \right) \mbox{ a.u.}
\label{hs3}
\end{eqnarray}
If we neglect the d-wave terms and use the experimental values \eref{1a}
for $A$ and $A_{\rm d}$, we can solve \eref{hs1} --~\eref{hs3}
for $x_{\rm s}$, $x_{\rm p}$ and $W_d$:
\begin{eqnarray}
        x_{\rm s} = 0.803, \qquad x_{\rm p} = 0.492,
\label{hs4a}\\
        W_d = -11.7 \mbox{ a.u.}= -1.45 \cdot 10^{25} \mbox{Hz/(e cm)},
\label{hs4}
\end{eqnarray}
where sings of $x_{\rm s}$ and $x_{\rm p}$ are chosen to account for the
repulsion of the unpaired electron from the F$^-$ ion.

Result \eref{hs4} is more accurate than one can expect from
\eref{hs1} --~\eref{hs3}. It follows from the proportionality between $W_d$
and $\sqrt{A A_{\rm d}}$:
\begin{eqnarray}
        W_d = -1.69 \cdot 10^{16} \sqrt{A A_{\rm d}}\mbox{ 1/(e cm)},
\label{hs}
\end{eqnarray}
which is based on the behaviour of s- and p-waves in the vicinity of the
nucleus where molecular potential is close to that of the nucleus. In
contrast to that, results \eref{hs4a} depend on the assumption that
spherical waves in \eref{4} correspond to the orbitals of Yb$^+$.

Below we calculate several corrections to \eref{hs4} and \eref{hs}, which
mostly account for contribution of the higher spherical waves.
If we use the normalization condition $x_{\rm s}^2+x_{\rm p}^2+x_{\rm d}^2
\approx 1$, we can solve equations \eref{hs1} --~\eref{hs3} with the
d-wave included:
\begin{eqnarray}
        x_{\rm s} = 0.803; \quad x_{\rm p} = 0.481,
        \quad x_{\rm d}= 0.356, \quad W_d = -11.2 \mbox{ a.u.}
\label{hs5}
\end{eqnarray}

Comparison of \eref{hs4} and \eref{hs5} gives the following correction
coefficient for the d-wave contribution to $W_d$:
\begin{equation}
        k_{\rm d} = 0.96.
\label{hs6}
\end{equation}
Note that the normalization condition give the upper bound for the
coefficient $x_{\rm d}$. So, it is likely, that \eref{hs5} and \eref{hs6}
somewhat overestimate the d-wave contribution.

\subsection*{Spin-doubling constant}

Let us estimate the mixing coefficient $c_{\rm p}$ in \eref{7a}:
\begin{eqnarray}
        c_{\rm p} &=& \frac{\langle \Pi(A_1)| H_{\rm so}
                      | \Sigma(X)\rangle}{E_X-E_A}
        = \frac{\langle \pi_{\rm p},\omega | H_{\rm so}
                      | \sigma_{\rm s},\omega \rangle}{E_X-E_A}
        \approx \frac{x_{\rm p} a_{\rm p} \xi_{6,1}}{\sqrt{2}(E_X-E_A)},
\label{10}
\end{eqnarray}
where we use \eref{4} and \eref{6} to calculate the numerator:
\begin{eqnarray}
        \langle \pi_{\rm p},\omega | H_{\rm so}
        | \sigma_{\rm s},\omega \rangle \approx x_{\rm p} a_{\rm p}
        \langle \tilde{6}p_1,-\case{1}{2}| H_{\rm so}
        | \tilde{6}p_0,\case{1}{2} \rangle \approx x_{\rm p} a_{\rm p}
        \langle 6p_1,-\case{1}{2}| H_{\rm so}
        |6p_0,\case{1}{2} \rangle,
\label{10a} \\
        \langle n,l,1,-\case{1}{2}| H_{\rm so}
        |n,l,0,\case{1}{2} \rangle \equiv
        -\xi_{n,l} \langle l,1|l_1|l,0 \rangle
        \langle-\case{1}{2}|s_{-1}|\case{1}{2}\rangle
        = \case{1}{2}\xi_{n,l}\sqrt{l(l+1)},
\label{10b}
\end{eqnarray}
and $\xi_{n,l}$ is the atomic spin-orbit constant for the
$(n,l)$-shell. It is proportional to the radial integral which enters
\eref{17c}. For the Yb$^+$ ion $\xi_{6,1}= 1900 \mbox{ cm}^{-1}$.

The level \eref{6} is identified as A$_1$ ($E_{A_1}$=18090~cm$^{-1}$).
It's fine splitting with the level A$_2$ ($\Pi_{3/2}$) is 1370~cm$^{-1}$
\cite{HH}.  Within the same approximation we can link this splitting to the
constant $\xi_{6,1}$ and find parameter $a_{\rm p}$:
\begin{eqnarray}
        E_{A_2}-E_{A_1} \approx a_{\rm p}^2 \xi_{6,1},
        \quad \Rightarrow \quad a_{\rm p}^2 \approx 0.72.
\label{9}
\end{eqnarray}

In a same manner we can calculate coefficient $c_{\rm f}$ in
\eref{7a}.  In this case the spin-orbit interaction mixes $\sigma_{\rm
f}$-hole with $\pi_{\rm f}$-hole:
\begin{eqnarray}
        c_{\rm f} &=& \frac{\langle \Pi(F)| H_{\rm so}
        | \Sigma(X)\rangle}{E_X-E_{F}} \approx \delta_{\rm f}
        \frac{\overline{\langle\pi_{\rm f},-\omega|} H_{\rm so}
        \overline{|\sigma_{\rm f},-\omega\rangle}}{E_X-E_{F}} \approx
        \frac{-\sqrt{3}\delta_{\rm f}\overline{\xi}_{4,3}}{E_X-E_{F}}.
\label{11}
\end{eqnarray}
The only problem here is that molecular state with the $\pi_{\rm f}$-hole,
which we define as $\Pi(F)$, is not known, and thus we do not know the
energy denominator.  The spin-orbit constant for the f-hole has the
opposite to normal sign:  $\overline{\xi}_{4,3}=-3665$~cm$^{-1}$.

Formulae \eref{10}, \eref{9} and \eref{11} reduce the number of
independent parameters in the wave function \eref{7a}.  Still, it has two
extra parameters $\delta_{\rm f}$ and $\varepsilon_{\rm d}$ as compared to
the wave function \eref{4} which was used in  \eref{hs1} --~\eref{hs3}.
Below we eliminate parameter $\delta_{\rm f}$ using experimental value of
the spin-doubling constant $\gamma$.

The spin-doubling term in the effective Hamiltonian \eref{1} arise from
the spin-orbit interaction. It was shown in \cite{KLM} that
\begin{eqnarray}
        \gamma = 2B \left(1 - \langle \Sigma(X),\case12|J_{e,+}
                |\Sigma(X),-\case12 \rangle \right),
\label{12}
\end{eqnarray}
where ${\bf J}_e={\bf L}+{\bf S}$ is the total angular momentum of the
electrons.  For the pure $\Sigma$ state $\langle{\bf J}_e \rangle =
\langle{\bf S}\rangle$ and \eref{12} gives $\gamma = 0$. This is no longer
true, when the spin-orbit corrections are taken into account.
With the help of \eref{12}, \eref{10} and \eref{11} it is
easy to calculate $\gamma$ for the state \eref{7a}:
\begin{eqnarray}
      \gamma = 2B \left(2(x_{\rm p} a_{\rm p})^2 \frac{\xi_{6,1}}{E_X-E_A}
      +12 \delta_{\rm f}^2\frac{\overline{\xi}_{4,3}}{E_X-E_F}\right).
\label{15}
\end{eqnarray}
Note that numerical factors in parentheses are equal to $l(l+1)$.

As we already pointed out, the second denominator in \eref{15} is
unknown. If we use corresponding energy interval for the Yb$^+$ ion, we
receive the following relation between $x_{\rm p}$ and $\delta_{\rm f}$
\begin{eqnarray}
        2100 x_{\rm p}^2 - 11150 \delta_{\rm f}^2 = 13,
        \quad \Rightarrow \quad
        \delta_{\rm f}^2 \approx 0.19 x_{\rm p}^2.
\label{17}
\end{eqnarray}
This equation shows that experimental value of $\gamma$ correspond to
almost complete cancelation between p-wave and f-wave contributions.
To obtain this relation between $\delta_{\rm f}$ and $x_{\rm p}$ we used
ionic spin-orbit constants $\xi_{n,l}$ and ionic denominator in
\eref{15}.  So, we can not expect it to be much better than an order
of magnitude estimate.

\subsection*{Spin-orbit and f-hole corrections to $W_d$}

It follows from \eref{17c}, \eref{15} and \eref{17}, that for
the molecular state \eref{7a} the f-wave contribution to the constant
$A_{\rm d}$ should be approximately 6 times smaller than that of the
p-wave.  Indeed, the two contributions to the constant $\gamma$ cancel
each other, but in \eref{15} there is the factor $l(l+1)$ which is not
present in \eref{17c}.

A straightforward relativistic calculation result in the following f-hole
correction to the hyperfine tensor
\begin{eqnarray}
     \delta A &=& -\frac{g_{\rm n} \alpha}{2m_{\rm p}} \delta_{\rm f}^2
                    \left(\frac{12}{49} h_{5/2,5/2}
                  + \frac{32}{49} h_{5/2,7/2}
                  + \frac{64}{147} h_{7/2,7/2}\right) \mbox{ a.u.},
\label{18} \\
     \delta A_{\rm d} &=& \frac{g_{\rm n} \alpha}{2m_{\rm p}}
                    \delta_{\rm f}^2
                    \left(\frac{48}{245} h_{5/2,5/2}
                  - \frac{8}{49} h_{5/2,7/2}
                  +\frac{64}{441} h_{7/2,7/2}\right) \mbox{ a.u.}
\label{19}
\end{eqnarray}
The radial integrals $h_{j,j'}$ here have the form
$
              h_{j,j'} = \int_0^{\infty}
              {{\rm d}r \left( f_j g_{j'}+g_j f_{j'} \right)},
$
where $f_j$ and $g_j$ are the upper and lower components of the radial
Dirac wave function. Again we take radial integrals for the Yb$^+$ ion and
use \eref{17} to arrive at
\begin{eqnarray}
     \delta A = -11 x_{\rm p}^2 \mbox{ MHz}, \qquad
     \delta A_{\rm d} = 64 x_{\rm p}^2\mbox{ MHz}.
\label{22}
\end{eqnarray}

It is clear, that correction \eref{22} to isotropic constant $A$ \eref{hs1}
is negligible, while correction to the dipole constant $A_{\rm d}$
\eref{hs2} is about 15\% for a given $x_{\rm p}$, which is in a good
agreement with our nonrelativistic estimate $\case{1}{6}$.

With the f-hole correction \eref{22} included, equation \eref{hs2} is
changed to
\begin{eqnarray}
    A_{\rm d} = \left(486 x_{\rm p}^2 + 36 x_{\rm d}^2 \right) \mbox{ MHz},
\label{23}
\end{eqnarray}
and we obtain the following f-hole correction coefficient to $W_d$
\begin{eqnarray}
     k_{\rm f} = 0.93.
\label{24}
\end{eqnarray}
In deriving \eref{24} we have ignored the direct f-wave contribution
to the constant $W_d$ (see \eref{hs3}). Such contribution is
proportional to a small product $x_{\rm d} \delta_{\rm f} \leq 0.1$ and
is less than 1\%.

The last correction to the constant $W_d$ is associated with the
admixture of the $\pi_{\rm p}$ state to \eref{7a}. According to \eref{10}
$c_{\rm p} \approx -0.06 x_{\rm p}$. This admixture changes weights of the
p$_{1/2}$- and p$_{3/2}$-waves in the wave function. That, in turn,
slightly changes coefficients in \eref{hs2} and \eref{hs3} (note, that
only p$_{1/2}$-wave contributes to $W_d$).  A simple calculation give
\begin{eqnarray}
     k_{\rm so} = 0.98.
\label{25}
\end{eqnarray}

\subsection*{Discussion}

Taking into account \eref{hs6}, \eref{24} and
\eref{25}, we obtain the total correction factor for $W_d$ to be
$k_{\rm corr} =k_{\rm d} k_{\rm f} k_{\rm so} = 0.87$. Applying it to
\eref{hs}, we arrive at
\begin{eqnarray}
     W_d = -10.2 \mbox{ a.u.}= -1.26 \cdot 10^{25} \mbox{Hz/(e cm)}.
\label{27}
\end{eqnarray}

We have calculated corrections to the semiempirical value of the
constant $W_d$ of the spin-rotational Hamiltonian \eref{1} which
can account for the $P,T$-odd effects in YbF molecule. The main
correction is caused by the f-hole admixture to the ground state. Two
other corrections account for the spin-orbit interaction and for the
d-wave term in the wave function. Altogether they reduce the answer by
more than 10\%.

We think that wave function \eref{7a} includes two most important
configurations. Admixture of other configurations will tend to decrease
constant $W_d$, so we can expect that our value \eref{27} is slightly
overestimated.  But it is unlikely, that corresponding corrections are
significantly larger than those discussed above. So, we estimate the
accuracy of our calculation to be about 20\%, which is typical to the
semiempirical method.

{\it Ab initio} calculation \cite{TME} gave smaller absolute value for
$W_d$, but it also underestimated both $A$ and $A_{\rm d}$. It is
more informative to compare the coefficient in the equation \eref{hs}
which correspond to different calculations. Our final value for this
coefficient is $-1.47 \cdot 10^{16}$~1/(e~cm), while results of \cite{TME}
correspond to $-1.75 \cdot 10^{16}$~1/(e~cm). So, in this sense, our result
is even smaller than that of \cite{TME}.

In this paper we have not considered other possible sources of
$P,T$-violation, such as scalar neutral currents \cite{GLM} and
magnetic quadrupole moment of the nucleus \cite{SFK}. Corresponding
constants $W_{\rm S}$ and $W_{\rm M}$ of the spin-rotational
Hamiltonian can be calculated in a similar way to $W_d$ \cite{KE}. Our
value for the former constant is: $W_{\rm S}=-43$~kHz.  Because of the
higher multipolarity of the electronic operator associated with the
constant $W_{\rm M}$, this interaction is much more sensitive to the
higher terms of the spherical wave expansion of the molecular wave
function. Thus, it is more difficult to make accurate calculation of
this interaction within the semiempirical approach.

\subsection*{Acknowledgments}

The author is grateful to E A Hinds, I B Khriplovich,
H M Quiney and A V Titov for valuable discussions.


\begin{thebibliography}{99}
\bibitem{KL}
        Kozlov M G and Labzowsky L N 1995 {\it J. Phys. B} {\bf 28} 1933
\bibitem{SPRH}
        Sauer B E, Peck S K, Redgrave G and Hinds E A 1996
        Private communication
\bibitem{K85}
        Kozlov M G 1985 {\it JETP} {\bf 62} 1114
\bibitem{KE}
        Kozlov M G and Ezhov V F 1994 \PR A {\bf 49} 4502
\bibitem{IBK}
        Khriplovich I B 1997 Private communication; quoted in
        {\sl CP-Violation Without Strangeness}, (Springer, in press)
\bibitem{SF}
        Sushkov O P, Flambaum V V 1978 {\it JETP} {\bf 48} 608
\bibitem{GLM}
        Gorshkov V G, Labzowsky L N and Moskalev A N
        1979 {\it ZhETF} {\bf 76} 414 [{\it JETP} {\bf 49}]
\bibitem{FK}
        Flambaum V V and Khriplovich I B 1985 \PL {\bf 110A} 12
\bibitem{KLM}
        Kozlov M G, Labzowsky L N and Mitruschenkov A O 1991
        {\it JETP} {\bf 73} 415
\bibitem{HH}
        Huber K P, Herzberg G 1979 {\it Molecular Spectra and
        Molecula Structure. IV. Constants of Diatomic Molecules}
        (Van Nostrand, New York)
\bibitem{SWH}
        Sauer B, Wang J, and Hinds E A 1995
        \PRL {\bf 74} 1554
\bibitem{KW}
        Knight L B, Jr. and Weltner W, Jr. 1970 {\it J. of Chem. Phys.}
        {\bf 53} 4111
\bibitem{TME}
        Titov A V, Mosiagin N S, Ezhov V F 1996
        \PRL {\bf 77} 5346
\bibitem{Q}
        Quiney H M, Skaane H and Grant I P 1997
        submitted to {\it Adv. Quant. Chem.}
\bibitem{PGHS}
        Sandars P G H 1965 \PL {\bf 14} 194; 1966
         {\bf 22} 290
\bibitem{IBK1}
        Khriplovich I B 1991 {\it Parity Non-Conservation in Atomic
        Phenomena} (Gordon and Breach, New York)
\bibitem{SFK}
        Sushkov O P, Flambaum V V and Khriplovich I B 1984
        {\it JETP} {\bf 60} 873
\end{thebibliography}
\end{document}